# On the concept of switching nonlinearity
## (a comment on "*Switching control of linear systems for generating chaos*" by X. Liu, K-L. Teo, H. Zhang and G. Chen)

*Emanuel Gluskin*

Holon Institute of Technology, Holon 58102, Israel, and Electrical Engineering Department, Ben-Gurion University of the Negev, Beer Sheva 84105, Israel.
gluskin@ee.bgu.ac.il
http://www.hit.ac.il/departments/electronics/staff/gluskin.htm

**Abstract**: It is explained and stressed that the chaotic states in [1] are obtained by means of *nonlinear switching*.

In the very interesting work [1] some chaotic states are obtained, but at no place nonlinearity of the system (or at all the term 'nonlinear') is mentioned, and the whole terminology of [1] can create impression that *by means of switching in a linear system* one can obtain chaos. However, it is well known that chaos can be obtained only in nonlinear systems, and it has to be clearly seen that in the method of [1] nonlinearity presents.

As is shown in [2] (see also [3] and [4]) *if the switching instants are defined by the unknowns to be found (i.e. the state-variables of the system under study), then the switched system is nonlinear*.

As is seen from Section 2 of [1], the *switching* there, from one linear system to another, is done at time instants associated with some constraints on the vector X($t$) of the state-variables of the system. (We use notations of [1].) In other words, these instants depend on X, $t_{1(2)} = t_{1(2)}(X)$. Thus, even though the switching is done in [1] between some per se linear systems, in view of [2] such a switching means *nonlinearity of the whole system*.

Since the switching instants that influence the matrix via the elements being interplaced (or, equivalently, one being changed), depend on X, such *whole system* obeys equations of the type

$$dX/dt = [A(t,X)]X + \ldots . \qquad (1)$$

Thus, it should be spoken in [1] about *nonlinear switching (control) of linear systems*.

It is important to see that if the switching instants were *prescribed*, -- the whole system would be *linear* (*time-variant* because of the switchings), satisfying the equations of the type

$$dX/dt = [A(t)]X + \ldots, \qquad (2)$$

and no chaos would be obtained.

We do think that the point of terminology is a part of correct academic outlook on the very complicated field of switched circuits.  There are so many chaotic circuits that, otherwise, in a particular research it indeed can be not easy for us "to see the forest behind the trees".

Certainly, it has to be clear that "switching between linear systems" can mean nonlinearity.

Hoping that this comment will be useful for the Readers of [1], we would like to stress that it is not purposed, in any sense, to decrease the importance of the method of chaos generation suggested in [1].  Moreover, the direction of seeking *complicated constraints* on X($t$) (in [1], using norms), and not only the simplest concept of zero-crossing, or level-crossing of some $x(t) \in$ X($t$), as in [2], in order to obtain some nonlinearity needed for chaos, seems to us to be promising.  In particular, such studies will better outline the borders of the concept of "switching nonlinearity" the usefulness of which is stressed in [2-4] and here.